\documentclass[pre,floatfix,aps,twocolumn,showpacs,superscriptaddresss]{revtex4-1}

\usepackage{graphicx}

\begin{document}

\title{Quantum Potts Models on the Sierpi\'nski Pyramid}

\author{Roman Kr\v cm\'ar$^1$}
\author{M\'{a}ria Zelenayov\'{a}$^{1,2}$}
\author{Jozef Genzor$^1$}
\author{Libor~Caha$^1$}
\author{Peter Rap\v can$^1$}
\author{Tomotoshi Nishino$^3$}
\author{Andrej Gendiar$^1$}
\affiliation{$^1$Institute of Physics, Slovak Academy of Sciences, D\'ubravsk\'a cesta 9, SK-845 11 
Bratislava, Slovakia}
\affiliation{$^2$Department of Theoretical Physics, Comenius University, Mlynsk\'a Dolina F2, SK-842 48 
Bratislava, Slovakia}
\affiliation{$^3$Department of Physics, Graduate School of Science, Kobe University, Kobe 657-8501, 
Japan}

\date{\today}

\begin{abstract}
Phase transition of the two- and three-state quantum Potts models on the Sierpi\'nski pyramid are 
studied by means of a tensor network framework, the higher-order tensor renormalization group method. 
Critical values of the transverse magnetic field and the magnetic exponent $\beta$ are evaluated. 
Despite the fact that the Hausdorff dimension of the Sierpi\'nski pyramid is exactly two $( = \log_2^{~} 4)$, 
the obtained critical properties shows that the effective dimension is lower than two.
\end{abstract}

\maketitle

\section{Introduction}

Dimensionality plays an important role in quantum phenomena. In one dimension, Luttinger liquid
behavior is common to a variety of quantum systems~\cite{Tomonaga,Luttinger}. In two dimensions, 
topological phases can play another important role, as has been widely studied for the electronic structure 
of Graphene and related materials. In phase transitions and critical phenomena, the lattice 
dimension is a key component for the universality~\cite{Domb}, in addition to a symmetry in the
local degrees of freedom and interaction range. For the classification of criticality, quantum models, 
such as the transverse-field Ising model, have been extensively studied on regular lattices in various 
spatial dimensions, including hyperbolic lattices which have infinite effective dimension~\cite{infinite}.

Fractal lattices can bring novel views on dimensional studies in critical phenomena since their Hausdorff 
dimensions can be non-integer. Quantum systems with fractal geometry have been occasionally studied,
partially because they appear in a variety of physical phenomena. For example, fractal nature 
arises when a quantum phase transition occurs in solids~\cite{richardella}. 
%
%
Besides, fractal structures have been inspiring scientist from various fields of physics.
Amongst them all, we mention that it is even possible to fabricate artificial fractals, such as
Sierpi\'nski triangle, on a solid surface~\cite{kempkes}. One also encounters presence of fractals
in quantum gravity~\cite{benedetti}.

\begin{figure}[tb!]
\includegraphics[width = 0.36 \textwidth]{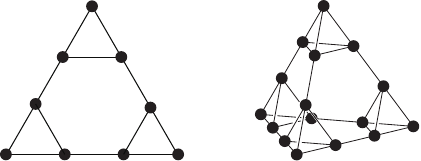}
\caption{Sierpi\'nski triangle (left) and the pyramid (right). 
We consider the fractal structures where adjacent units are connected by a bond.}
\label{frc}
\end{figure}

In this article we focus on quantum lattice models on the Sierpi\'nski fractals. Figure~\ref{frc} shows small 
finite size lattices with the geometry of the Sierpi\'nski triangle (left) and the pyramid (right). Both of 
these lattices can be constructed recursively. In the case of the Sierpi\'nski triangle, an elementary 
unit consists of $3$ lattice sites that are located at the vertices of a triangle, and the extension of the 
system is performed by connecting $3$ units so that the adjacent units are connected through an additional 
bond. In the case of Sierpi\'nski pyramid, the elementary unit is a tetrahedron that consists of $4$ lattice 
sites, and $4$ units are connected through $6$ additional bonds. Repeating such extension process recursively, 
one can construct a fractal lattice of an arbitrary size. The coordination number is $3$ in the case of the
Sierpi\'nski triangle thus created, and is $4$ for the Sierpi\'nski pyramid. 

There is another type of Sierpi\'nski triangle and pyramid, where elementary units are joined so that 
each site is shared by adjacent units. Under this extension scheme, the coordination number of the 
Sierpi\'nski triangle is $4$, and that of the pyramid is $6$. 
In order to avoid any confusion, we refer to these ``site-sharing fractals'' as the B-type,
and the ``bond-sharing fractals'' shown in Fig.~~\ref{frc} as the A-type in the following.
It should be noted that the fractal dimension in the thermodynamic (i.e., large system-size) limit
does not depend on the choice of the A- or B-type.

As a typical example of quantum lattice models, the $q$-state Potts models, which include the 
transverse-field Ising (TFI) model as the case $q = 2$, have been studied on the Sierpi\'nski triangle.
A method of analysis is the conventional real-space renormalization group (RSRG). 
Kubica and Yoshida obtained the critical indices $\nu = 0.7196$ and $\nu = 0.6213$, respectively, for 
the cases $q = 2$ and $q = 3$~\cite{Kubica}. The same result was also reported by Xu {\it et al.}~\cite{Xu}. 
Another choice of the numerical method is the quantum Monte-Carlo (QMC) simulation, where Yi obtained 
$\nu = 0.66 \pm 0.05$ for the case $q = 2$~\cite{Yi_1}. They also reported $\nu = 0.53 \pm 0.03$ 
for the case $q = 3$~\cite{Yi_2}. Independently, Yoshida and Kubica reported $\nu = 0.76 \pm 0.01$ 
for $q = 2$~\cite{Yoshida}. The Table~I summarizes these results, including the recent study performed by 
the higher-order tensor renormalization group (HOTRG) method~\cite{HOTRG}. 

\begin{table}
\caption{Critical exponents of the quantum Potts models ($q = 2$ and $q = 3$) on the Sierpi\'nski triangle.
The lattice A- or B-type are shown as the superscripts of the method used to avoid any other confusions.}
\begin{tabular}{cllll}
\hline
\hline
~~$q$~~ &  ~~Method & ~~~~~~$\nu$ & ~~~~~$\beta$ & ~~~$\gamma$  \\
\hline
2 & RSRG$^{\rm A}_{~}$~\cite{Kubica,Xu} & ~~$0.7196$ & &  \\
 & QMC$^{\rm B}_{~}$~\cite{Yi_1} & ~~$0.66(5)$ & ~~$0.19(2)$ & ~$1.45(5)$  \\
 & QMC$^{\rm A}_{~}$~\cite{Yoshida} & ~~$0.76(1)$ & ~~$0.18$ & ~$1.60$  \\
& HOTRG$^{\rm A}_{~}$~\cite{roman} & ~~& ~~$0.20$ &   \\
\hline
3 & RSRG$^{\rm A}_{~}$~\cite{Kubica,Xu} & ~~$0.6213$ & &  \\
 & QMC$^{\rm B}_{~}$~\cite{Yi_2} & ~~$0.53(3)$ & ~~$0.145(10)$ & ~$1.24(3)$  \\
\hline
\hline
\end{tabular}
\end{table}

Relatively less is known on the Sierpi\'nski pyramid. By means of RSRG, the exponents 
$\nu = 0.6174$ and $\nu = 0.5390$ are known, respectively, for the case $q = 2$ and $q = 3$~\cite{Kubica,Xu}.
By QMC simulations, Yi reported $\nu = 0.62 \pm 0.05$~\cite{Yi_1} and 
$\nu = 0.43 \pm 0.02$~\cite{Yi_2}, respectively, for the cases $q = 2$ and $q = 3$. Independently, 
Yoshida and Kubica obtained $\nu = 0.660 \pm 0.005$ for $q = 2$, and they conjectured that the 
quantum $q = 3$ Potts model exhibits a discontinuous (first-order) phase transition on the Sierpi\'nski
pyramid~\cite{Yoshida}. These results for the pyramid are summarized in Table~II. 

\begin{table}
\caption{Critical exponents of the quantum Potts models ($q = 2$ and $q = 3$) on the Sierpi\'nski pyramid,
including $\beta$ we have calculated by means of the HOTRG method in this work.}
\begin{tabular}{cllll}
\hline
\hline
~~$q$~~ &  Method & ~~~~$\nu$ & ~~~$\beta$ & ~~~$\gamma$  \\
\hline
2 & RSRG$^{\rm A}_{~}$~\cite{Kubica} & ~$0.6174$ & & \\
 & QMC$^{\rm B}_{~}$~\cite{Yi_1} & ~$0.62(5)$ & ~$0.25(2)$ & ~$1.55(5)$ \\
 & QMC$^{\rm A}_{~}$~\cite{Yoshida} & ~$0.66(5)$ & & \\
 & HOTRG$^{\rm A}_{~}$ & & ~$0.232$ & \\
\hline
3 & RSRG$^{\rm A}_{~}$~\cite{Kubica}  & ~$0.5390$ & & \\
 & QMC$^{\rm B}_{~}$~\cite{Yi_2} & ~$0.43(2)$ & ~$0.15(1)$ & ~$1.18(5)$ \\
 & HOTRG$^{\rm A}_{~}$ & & ~$0.154$ & \\
\hline
\hline
\end{tabular}
\end{table}

Along with these studies of phase transitions on Sierpi\'nski fractals, it turned out that the
definition of the {\it effective dimension} is not straightforward. The calculated exponents do
not agree with the hyper-scaling hypothesis if the Hausdorff dimension of the fractal lattice
is considered as the effective spatial dimension. A similar discrepancy is also reported for the
classical Ising model defined on fractal lattices~\cite{jozef_1,jozef_2}.
For the purpose of getting better insight into the effective dimension, we perform a precise numerical 
study for the $q = 3$ quantum Potts model on the Sierpi\'nski triangle and pyramid by means of the HOTRG 
method. 

The structure of this article is as follows: In the next section we introduce the Potts models on
the Sierpi\'nski pyramid. In Section III we show the numerical results. Conclusions are summarized
in the last section.

\section{Quantum Potts model}

The quantum $q$-state Potts model is described by the lattice Hamiltonian
\begin{equation}
{\hat H} = - J \sum_{\langle i, j \rangle}^{~} \delta\left( {\hat s}_i^{~}, {\hat s}_j^{~} \right) - 
h \sum_{i}^{~} \, \sum_{k = 1}^{q - 1} \left( {\hat \Gamma}_i^{~} \right)^k_{~} \, ,
\label{P-Ham}
\end{equation}
where ${\hat s}_i^{~}$ is the diagonal operator, whose eigenvalues are integers from $1$ 
to $q$, on the lattice site labeled by $i$~\cite{Yi_1,Yi_2,ding}. The summation of the first term
on the r.h.s. runs over pairs of neighboring sites, it is denoted by $\langle i, j \rangle$,
and $J$ parameterizes the ferromagnetic interaction. We assume that the system is on
a sufficiently large Sierpi\'nski pyramid. The second term represents the quantum flipping 
effect by means of the transverse field, which is parameterized by the constant magnetic field $h$.
The matrix representation of the operator ${\hat \Gamma}_i^{~}$ is nothing but the shift matrix
\begin{equation}
\Gamma_i^{~} = \left( \begin{array}{cc} 0 & I_{q - 1} \\ 1 & 0 \end{array} \right),
\end{equation}
where $I_{q - 1}$ is an identity matrix of the dimension $q - 1$. In the case of $q = 2$, the Hamiltonian
${\hat H}$ in Eq.~(\ref{P-Ham}) coincides with the transverse-field Ising model if rescaling $J \to 2J$.
In the following we consider the cases $q = 2$ and $3$ only. 
We focus on the ground-state phase transition of the system, which is located at a certain value
of the transverse field $h=h_{\rm c}^{~}$.

In order to analyze the ground-state properties of the system, we use the HOTRG method~\cite{HOTRG}.
As it has been done in the previous study on the Sierpi\'nski  triangle~\cite{roman},
we introduce the Trotter-Suzuki decomposition~\cite{Trotter,Suzuki_1,Suzuki_2} of
the imaginary-time path integral representation of the thermal density matrix.
The first and the second terms on the r.h.s. of Eq.~(\ref{P-Ham}) are treated
as a non-commuting pair of operators ${\hat H}_0^{~}$ and ${\hat H}_1^{~}$. 
Typically, we choose the imaginary time step $\Delta \tau = 0.01$. The classical lattice system obtained 
through this decomposition naturally has a network structure that consists of a local weight represented
by $6$-leg tensors, where two legs correspond to the imaginary-time degrees of freedom.
Since the tensor network is highly anisotropic,
a couple of tensors stacking along the imaginary time direction is grouped in advance~\cite{roman}
(we stacked $7$ tensors at most). Thus stacked tensors are then renormalized forming another tensor. After 
such pre-processing, we started the conventional HOTRG procedure~\cite{HOTRG}, which expands the corresponding 
tensors recursively by alternating the three space and one imaginary time directions.
Details of the numerical implementation can be found in Ref.~\cite{roman}. 

In order to detect the phase transition, we calculate the expectation value of local magnetization 
\begin{equation}
M = \frac{q}{q-1} \, 
%
%
\left[ \langle \delta\left( {\hat s}, 1 \right) \rangle - \frac{1}{q} \right]
\end{equation}
%
%
averaging inner spin operators ${\hat s}$ represented by impurity tensors~\cite{jozef_1}.
Such an observation of $M$ can be performed by keeping the renormalized expression of ${\hat s}$
in each of the renormalization-group (RG) transformations~\cite{HOTRG}. During the HOTRG calculations,
we kept $\chi = 40$ for the block-spin states at most.

\section{Numerical results}

\begin{figure}[tb]
\centering
\includegraphics[width = 0.46 \textwidth]{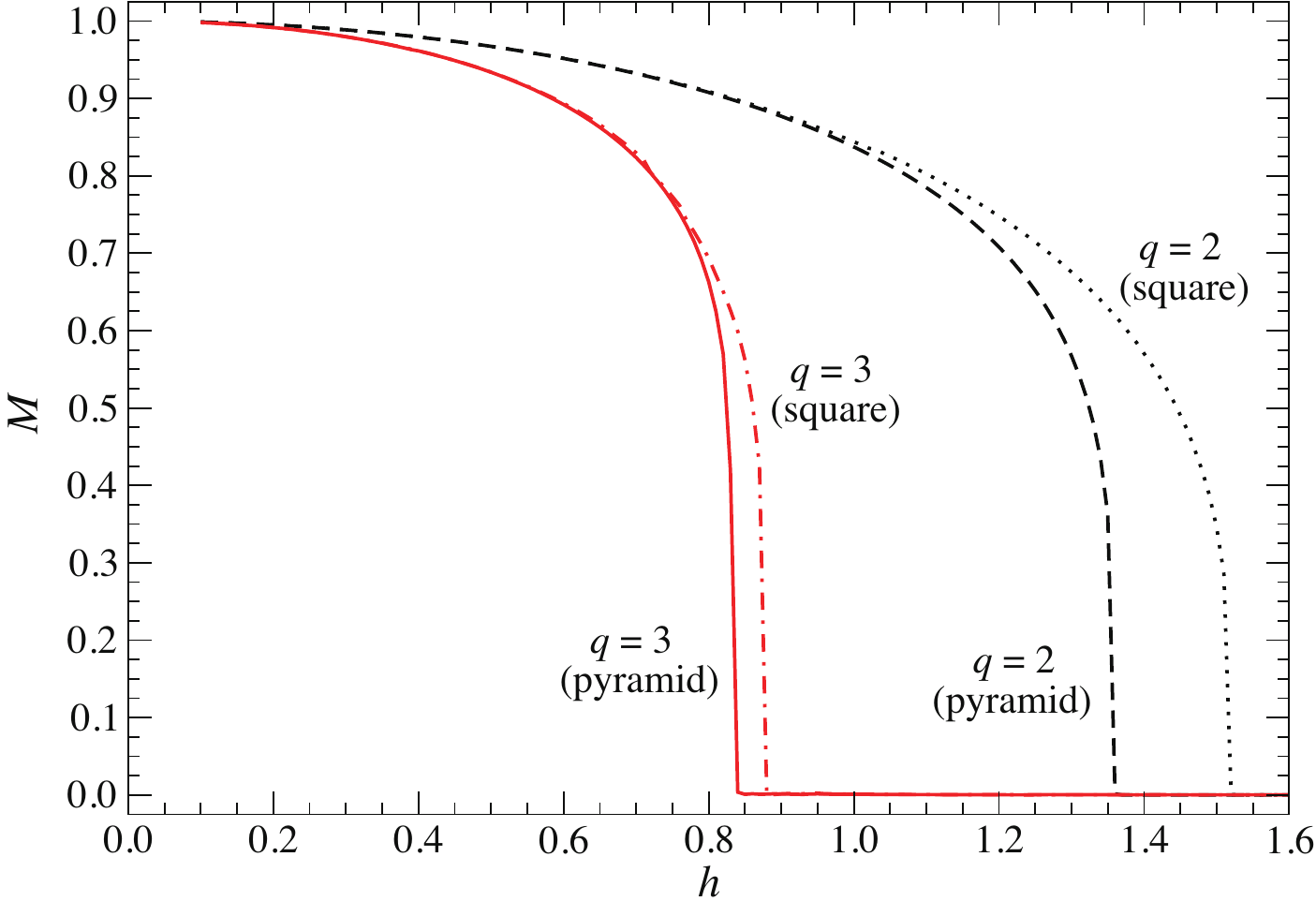}
\caption{Ground-state magnetization $M$ with respect to the transverse field $h$,
calculated for $q = 2$ and $q = 3$ Potts models on Sierpi\'nski pyramid.
For comparison, the square-lattice data are shown, too.}
\label{magnetization}
\end{figure}

Let us recall that the Sierpi\'nski pyramid (of the A-type) has the identical coordination number as
the square lattice. Figure~\ref{magnetization} shows the calculated ground-state magnetization $M$
with respect to the transverse field $h$ 
for $q = 2$ and $q = 3$ Potts models on the Sierpi\'nski pyramid. For comparison, the values of $M$ 
calculated on the square lattice are also plotted. In the small $h$-region, where $M$ is close to the unity,
the value of $M$ is insensitive to the global structure of the lattice. This is because when the correlation 
length is small, the effect of the coordination number is dominant. On the other hand, as $h$ increases, the
difference between the Sierpi\'nski pyramid and the square lattice becomes clearer.
The correlation length grows to infinity toward the critical field $h_{\rm c}^{~}$.

\begin{figure}[tb]
\centering
\includegraphics[width = 0.46 \textwidth]{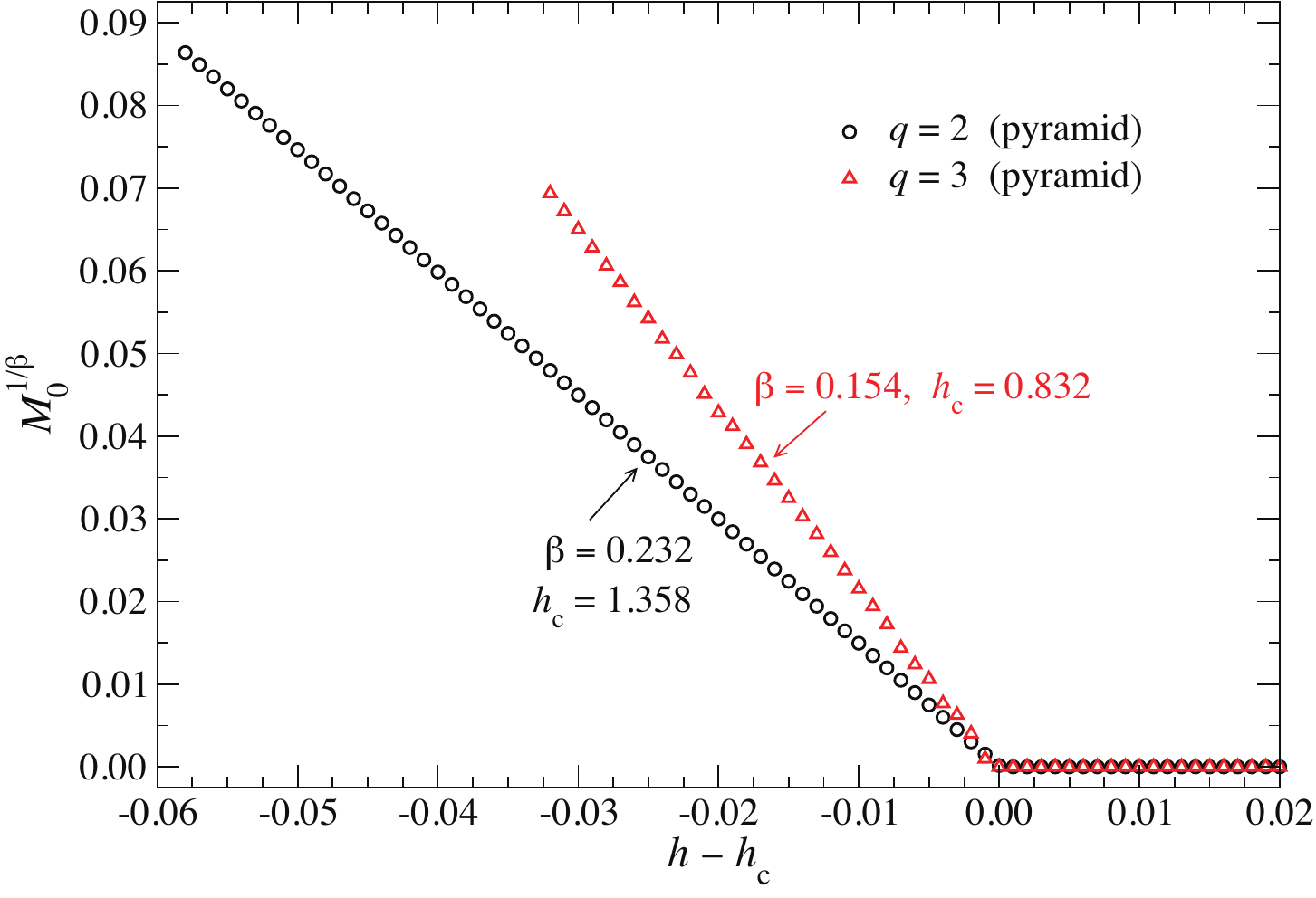}
\caption{The linear decay of the magnetization $M^{1/\beta}_{~}$, when $h - h_{\rm c}^{~} <0 $,
confirms the correctness of the obtained $\beta$ and $h_{\rm c}^{~}$.
For the case $q = 2$, $\beta = 0.232$ and $h_{\rm c}^{~} = 1.358$ are assumed.
For $q = 3$, we used $\beta = 0.154$ and $h_{\rm c}^{~} = 0.832$.}
\label{confirm}
\end{figure}

In order to analyze the critical singularity in the magnetization $M$, we evaluate the critical
exponent $\beta$ by fitting the calculated data with the scaling formula
\begin{equation}
M = C \left( h_{\rm c}^{~} - h \right)^{\beta}_{~}\, ,
\end{equation}
where $h_{\rm c}^{~}$ is the critical value of the transverse field, and $C$ is another fitting constant.
On the Sierpi\'nski pyramid, we obtained $\beta = 0.232$ and $h_{\rm c}^{~} = 1.358$ for the case 
$q = 2$ as a result of the fitting. For the case $q = 3$, we calculated $\beta = 0.154$ and
$h_{\rm c}^{~} = 0.832$. To confirm these estimated values, we plot $M^{1/\beta}_{~}$ in
Fig.~\ref{confirm}, assuming these obtained exponents. Evidently, below the critical field $h_{\rm c}^{~}$,
we observed a linear decrease of $M^{1/\beta}_{~}$ for both cases.

\section{Conclusions and discussions}

We have calculated the ground-state magnetization $M$ of the quantum Potts models on the 
Sierpi\'nski pyramid (of the A-type) for the cases $q = 2$ and $q = 3$. The estimated critical field 
$h_{\rm c}^{~}$ and the exponent $\beta$ are listed in Table~III, together with the related values
reported so far~\cite{HOTRG,Yi_1,Yi_2}. It is obvious that the value of $h_{\rm c}^{~}$ are consistent
with the Monte Carlo studies~\cite{Yi_1,Yi_2} on the B-type pyramid. This fact suggests that the
effective dimension is insensitive to the choice of lattice from A- or B-type.
The calculated $\beta$ shows that the effective dimension of the Sierpi\'nski pyramid is less
than two, when the critical universality is considered. This is in accordance with previous 
studies~\cite{Kubica,Xu,Yi_1,Yi_2,Yoshida}.

The calculation of thermodynamic functions, such as internal energy and entropy would provide further
information on the dimensionality and the scaling relation on the Sierpi\'nski pyramid~\cite{progress}.
In order to increase the variety of the fractal lattice structure is another direction of the future
study. For example, considering a higher-dimensional generalization of the Sierpi\'nski pyramid for 
the purpose of finding out the upper critical (spatial) dimension is a reachable study with
the use of the current computational resources. Evaluation of the entanglement entropy would be
a challenging task within the numerical tensor-network frameworks.

\begin{table}[tb]
\caption{List of critical field $h_{\rm c}^{~}$ and exponent $\beta$. 
}
\centering
\begin{tabular}{|l|l|l|l|l|}
\hline
\rule{0pt}{3.0ex} & \multicolumn{2}{|c|}{HOTRG$^{\rm A}_{~}$} & \multicolumn{2}{|c|}{MC$^{\rm B}_{~}$} \\
\cline{2-5}
\rule{0pt}{3.0ex} Model (lattice) & \quad\ \ $h_{\rm c}^{~}$ & \quad\ \ $\beta$ & \quad\ \ $h_{\rm c}^{~}$ & \quad\, \,$\beta$ \\
\hline 
\rule{0pt}{3.0ex} $q = 2$ (square) & 1.5219~\cite{HOTRG} & 0.3295~\cite{HOTRG} & 1.522~\cite{Yi_2} & 0.31~\cite{Yi_2}\\
\hline
\rule{0pt}{3.0ex} $q = 2$ (pyramid) & 1.358 & 0.232 & 1.3535~\cite{Yi_2} & 0.25~\cite{Yi_2}\\ 
\hline 
\rule{0pt}{3.0ex} $q = 3$ (square) & 0.876 & --- & 0.873~\cite{Yi_2} & --- \\ 
\hline
\rule{0pt}{3.0ex} $q = 3$ (pyramid) & 0.832 & 0.154 & 0.8207~\cite{Yi_2} & 0.15~\cite{Yi_2}\\
\hline
\end{tabular}
\end{table}

\section{Acknowledgment}

This work was supported by Agent\'{u}ra pre Podporu V\'{y}skumu a V\'{y}voja (No. APVV-20-0150), Vedeck\'{a} Grantov\'{a} Agent\'{u}ra M\v{S}VVaM SR and SAV (VEGA No. 2/0156/22). J.~G. is funded by the EU NextGenerationEU through the Recovery and Resilience Plan for Slovakia under the project No. 09I03-03-V04-00682. T.~N. and A.~G. acknowledge the support of Grant-in-Aid for Scientific Research.


\begin{thebibliography}{1}
\bibitem{Tomonaga} S.~Tomonaga, Prog. Theor. Phys. {\bf 5}, 554 (1950).
\bibitem{Luttinger} J.M.~Luttinger, J. Math. Phys. {\bf r}, 1154 (1963).
\bibitem{Domb} C.~Domb and M.S.~Green, ``phase transitions and Critical Phenomana'', (Academic Press, 1972).
\bibitem{infinite} A.~Gendiar, M.~Daniska, R.~Krcmar, and T.~Nishino, Phys. Rev. E {\bf 90}, 012122 (2014).
\bibitem{richardella} A.~Richardella, P.~Roushan, S.~Mack, B.~Zhou, D.A.~Huse, D.D.~Awschalom, A.~Yazdani, 
Science {\bf 327}, 665 (2010).
\bibitem{kempkes} S.N.~Kempkes, M.R.~Slot, S.E.~Freeney, S.J.M.~Zevenhuizen, D.~Vanmaekelbergh, 
\bibitem{benedetti} D.~Benedetti, Phys. Rev. Lett. {\bf 102}, 111303 (2009).
I.~Swart, and C.M.~Smith, Nature Physics {\bf 15}, 127 (2019).
\bibitem{Kubica} A.~Kubica and B.~Yoshida, arXiv:1402.0619. 
\bibitem{Xu} Y.L.~Xu, X.M.~Kong, Z.Q.~Liu, and C.C.~Yin {\bf 95}, 042327 (2017). 
\bibitem{Yi_1} H.~Yi, Phys. Rev. E {\bf 91}, 012118 (2015). 
\bibitem{Yi_2} H.~Yi, Phys. Rev. E {\bf 96}, 062105 (2017). 
\bibitem{Yoshida} B.~Yoshida and A.~Kubica, arXiv:1404.6311. 
\bibitem{HOTRG} Z.Y.~Xie, J.~Chen, M.P.~Qin, J.W.~Zhu, L.P.~Yang, and T.~Xiang, 
Phys. Rev. B {\bf 86}, 045139 (2012).
\bibitem{roman} R.~Krcmar, J.~Genzor, Y.~Lee, H.~Cencarikova, and T.~Nishino, 
Phys. Rev. E {\bf 98}, 062114 (2018). 
\bibitem{jozef_1} J.~Genzor, A.~Gendiar, and T.~Nishino, Phys. Rev. E {\bf 93}, 012141 (2016).
\bibitem{jozef_2} J.~Genzor, A.~Gendiar, and T.~Nishino, Phys. Rev. E {\bf 107}, 044108 (2023).
\bibitem{ding} C.~Ding, Y.~Wang, Y.~Deng, and H.~Shao, arXiv:1702.02675.
\bibitem{Trotter} H.F.~Trotter, Proc. Amer. Math. Soc. {\bf 10}, 545 (1959).
\bibitem{Suzuki_1} M.~Suzuki, J. Phys. Soc. Jpn. {\bf 21}, 2274 (1966).
\bibitem{Suzuki_2} M.~Suzuki, Prog. Theor. Phys. {\bf 56}, 1454 (1976).
\bibitem{progress} in progress.

\end{thebibliography}
\end{document}